\documentclass{article}

\usepackage{arxiv}
\usepackage{amsmath}
\usepackage{amsfonts}
\usepackage{amssymb}
\usepackage{lmodern}
\usepackage{subfiles}
\usepackage[T1]{fontenc}
\usepackage{mathrsfs}
\usepackage{bm}
\usepackage[colorlinks]{hyperref}

\usepackage{graphics}
\usepackage{graphicx}
\usepackage{epstopdf}
\usepackage{epsfig}
\usepackage{url}

\usepackage{subcaption}
\graphicspath{{./Figures/}}
\usepackage[font=small,labelfont=bf]{caption}
\newtheorem{remark}{Remark}
\newcommand{\R}{\mathbb{R}}
\newcommand{\C}{\mathbb{C}}
\newcommand{\N}{\mathbb{N}}
\newcommand{\V}{V\!a}

\usepackage[utf8]{inputenc} 
\usepackage[T1]{fontenc}    
\usepackage{booktabs}       
\usepackage{amsfonts}       

\title{Effect of Vadasz term on the onset of convection in a Darcy-Brinkman anisotropic rotating porous medium in LTNE}
\author{F. Capone \href{https://orcid.org/0000-0002-0672-999X}{\includegraphics[scale=0.1]{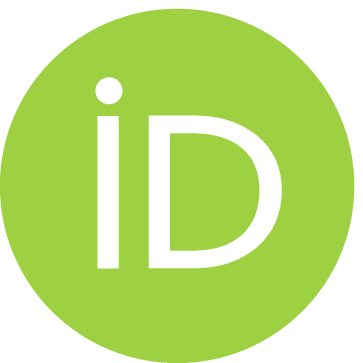}} \\ Dipartimento di Matematica e Applicazioni `R.Caccioppoli' \\ Universit\'a degli Studi di Napoli Federico II \\ Via Cintia, Monte S.Angelo, 80126 Napoli \\ Italy \\ \texttt{florinda.capone@unina.it} \\   \and J.A. Gianfrani\href{https://orcid.org/0000-0001-9906-2495}{\includegraphics[scale=0.1]{orcid.eps}} \\ Dipartimento di Matematica e Applicazioni `R.Caccioppoli' \\ Universit\'a degli Studi di Napoli Federico II \\ Via Cintia, Monte S.Angelo, 80126 Napoli \\ Italy \\ 
	\texttt{jacopoalfonso.gianfrani@unina.it}}
\begin{document}
\maketitle
\renewcommand{\shorttitle}{Effect of Vadasz term on the onset of convection}
\begin{abstract}
	In the present paper, the effect of the Vadasz inertia term on the onset of convective motions for a Darcy-Brinkman model is investigated. It is proved that this term leads to the possibility for oscillatory convection to occur. Hence, convection can occur via either oscillatory or steady motions. It is proved analytically that the onset of steady convection is not affected by the Vadasz term, while oscillatory convection is favoured by it. 
	Moreover, conditions to rule out the occurrence of oscillatory convection are determined numerically.
	
	The influence of rotation, interaction coefficient and mechanical and thermal anisotropies on the onset of instability is investigated, both analytically and numerically.
\end{abstract}

\section{Introduction}
Ever since Horton and Rogers \cite{Horton_Rogers} and Lapwood \cite{Lapwood} modelled for the first time the fluid motion in presence of a porous medium heated from below, the onset of natural thermal convection in porous media has been widely studied over the years(\cite{Arnone}, \cite{Massa}, \cite{Reddy}, \cite{Rees2022}). The several applications for this physical phenomenon makes this problem fascinating and of great interest. For instance, managing the heat transfer by the onset of convection is very important in industrial field, where metal foams, heat exchangers and systems of storage or removal of heat are employed (see \cite{Nield-Bejan}).

Given the main usage in industrial and engineering fields, porous media are usually man-made. Metal foams, for example, show very high porosity so that the fluid encounters few obstacles during its motion. As a result, to model such a porous medium, it is more convenient to employ the Brinkman law in place of the Darcy one (see \cite{caponeIJFM2016} \cite{Mal2004} \cite{Rees2001} \cite{Capone2012799} \cite{Yadav2020} \cite{Yadav2017} \cite{Barletta2011}). 

Moreover, in order to enhance or lower the heat transfer, porous media are built in such a way that they exhibit anisotropic behaviour in some of their features \cite{Gianfrani2020} \cite{Capone2009205} \cite{Tyv2015}. In this study, we considered a porous medium with horizontal isotropic permeability and thermal conductivity, following the definitions in \cite{Gov}, \cite{Cap2019}. The most simple example for this kind of porous medium is a sedimentary porous rock, which exhibits a layered configuration, favouring the fluid motion in the horizontal direction rather than the vertical one. 

In addition, in this study, the heat transfer through the porous medium is modelled under the assumption of Local Thermal Non Equilibrium (LTNE). According to this hypothesis, two different temperatures are defined: one for the fluid phase, one for the solid phase. In such a way, heat exchange between the phases is allowed. The assumption of LTNE plays an important role in physical industrial problems where quick heat transfer is involved (see \cite{Straughan2015book} for more details) and, more in general, when fluid thermal conductivity is much different from the solid one or when the fluid velocity is sufficiently high \cite{InghamPop2005} \cite{Rees2008}. Early studies on thermal convection in porous media in LTNE are \cite{Banu} and \cite{Straughan2006}, where both a linear and nonlinear analysis is carried out. More recently, the effect of LTNE on the onset of convection has been studied coupled to the effect of variable viscosity \cite{GianfraniActa} \cite{Shiv2010}, hyperbolic temperature equation \cite{Gianfrani2022} \cite{Eltayeb} \cite{Shiv}, symmetric wall heat flux \cite{Barletta_Kuz}, full anisotropy \cite{Siddabasappa} and high porosity  \cite{Freitas2022}.

The aim of this paper is to highlight physical and mathematical implications of the presence of the Vadasz inertia term in the momentum equation. In many industrial and engineering configurations the onset of convection in a rotating porous medium heated from below may occur exhibiting an oscillating behaviour in time. Such a behaviour is very useful in cooling systems when, for example, one is interested in a time modulated cooling down.
From a mathematical view point, in order to take into account oscillatory convective motions, the inertia term needs to be retained within the model. In 1998, Vadasz performed a remarkable study \cite{Vadasz1998} where he proved that the presence of inertia term in a model for a rotating porous medium leads to the occurrence of convection through oscillatory motions, which are not allowed when the inertia term is neglected. That paper is the reason why we refer to the inertia term in the momentum equation as the "Vadasz term".

The outline of the paper is the following:  Section \ref{vad_sec1} is devoted to the mathematical model describing the fluid motion in presence of a rotating anisotropic porous medium in LTNE. We determine the basic steady solution, whose stability we are interested in and consequently we obtain the dimensionless system of perturbations. In Section \ref{vad_sec2}, linear instability analysis is performed in order to determine the critical Rayleigh numbers for both steady and oscillatory convection. Finally, Section \ref{vad_sec3} involves the study of the effect of anisotropic permeability, anisotropic thermal conductivities, rotation, the Darcy number and, most importantly, the Vadasz number on the onset of convection.

\section{Mathematical model}\label{vad_sec1}
Let $\mathscr{F}$ be an incompressible fluid, initially at rest, saturating a horizontal porous layer, whose depth is $d$, confined between two planes, $z=0$ and $z=d$. We assume the layer to be heated from below and we denote by $T_L$ the temperature of the lower plane $z=0$ and by $T_U$ the temperature of the upper plane $z=d$. In addition, the layer rotates about the upward vertical axis $z$ with constant angular velocity $\Omega$. Therefore, the Coriolis force affects the fluid motion within the porous medium, together with gravitational and drag forces. Moreover, we assume that fluid and solid phases are not in thermal equilibrium, namely heat exchanges between the phases are allowed. Then, we refer to the fluid temperature with $T_f$ and to the solid temperature with $T_s$, by saying that the porous medium is in local thermal non-equilibrium (LTNE). Then
\begin{equation}
T_s=T_f=T_L \qquad \text{on} \ z=0, \qquad T_s=T_f=T_U \qquad \text{on} \ z=d
\end{equation}
with $T_L>T_U$.

In addition, the porous medium is horizontally isotropic, i.e. its features, such as thermal conductivity and permeability, are homogeneous in the horizontal direction. Let $\mathcal{K}_{*}$ be the permeability tensor, $\mathcal{D}^{*}_s$ and $\mathcal{D}^{*}_f$ be the thermal conductivity tensors of solid phase and fluid phase, respectively. By assuming that the principal axis $(x,y,z)$ of $\mathcal{K}_{*}$ coincide with the conductivity tensors' ones, it turns out (see \cite{Gov}, \cite{Cap2019})
\begin{equation}
\begin{aligned}
& \mathcal{K} = K_z \mathcal{K}_{*} \qquad \mathcal{K}_{*} = \begin{pmatrix}
\xi & 0 & 0 \\
0 & \xi & 0 \\
0 & 0 & 1
\end{pmatrix} \qquad
\xi = \frac{K_H}{K_z} \\
& \mathcal{D}_s = \kappa_z^s \mathcal{D}^{*}_s \qquad \mathcal{D}^{*}_s =  \begin{pmatrix}
\zeta & 0 & 0 \\
0 & \zeta & 0 \\
0 & 0 & 1
\end{pmatrix} \qquad
\zeta = \frac{\kappa_H^s}{\kappa_z^s} \\
& \mathcal{D}_f = \kappa_z^f \mathcal{D}^{*}_f \qquad \mathcal{D}^{*}_f =  \begin{pmatrix}
\eta & 0 & 0 \\
0 & \eta & 0 \\
0 & 0 & 1
\end{pmatrix}  \qquad
\eta = \frac{\kappa_H^f}{\kappa_z^f}
\end{aligned}
\end{equation}

Since the problem at stake involves a porous medium with high porosity, a more suitable Darcy-Brinkman model needs to be adopted (see \cite{Nield-Bejan}). Under these hypotheses, starting from models proposed in \cite{Gianfrani2021b}-\cite{Malashetty2010}, the mathematical model is
\begin{equation}\label{vad_mod}
\begin{cases}
& \rho_f c_a \textbf{v}_{,t} = -\nabla p - \dfrac{2 \Omega \rho_f}{\varepsilon} \textbf{k} \times \textbf{v} - \mu \mathcal{K}^{-1}\cdot \textbf{v}  + \rho_f g \alpha T_f \textbf{k} + \tilde{\mu} \Delta \textbf{v} \\
& \nabla \cdot \textbf{v} =0 \\
& \varepsilon (\rho c)_f  T^f_{,t} + (\rho c)_f \textbf{v} \cdot \nabla T_f = \varepsilon \nabla \cdot (\mathcal{D}_f \cdot \nabla T_f) + h(T_s - T_f) \\
& (1-\varepsilon)(\rho c)_s T^s_{,t} = (1-\varepsilon) \nabla \cdot (\mathcal{D}_s \cdot \nabla T_s ) - h(T_s - T_f)
\end{cases}
\end{equation}
where the Oberbeck-Boussinesq approximation is adopted and where $\textbf{v}$, $p$, $T_f$ and $T_s$ are (seepage) velocity, reduced pressure, fluid phase temperature and solid phase temperature, respectively; while $\mu$, $\tilde{\mu}$, $\rho_f$, $\rho_s$, $c$, $g$, $\alpha$, $c_a$, $\varepsilon$, $\Omega$, $h$ are dynamic and effective viscosity, fluid density, solid density, specific heat, gravity acceleration, thermal expansion coefficient, acceleration coefficient, porosity, angular velocity and interaction coefficient, respectively.

The boundary conditions (\ref{vad_BC1}) are coupled to (\ref{vad_mod})
\begin{equation}\label{vad_BC1}
\begin{aligned}
& T_s = T_f = T_L \quad \text{on} \ z=0, \qquad T_s = T_f = T_U \quad \text{on} \ z=d, \\
& \textbf{v} \cdot \textbf{n} = 0 \quad \text{on} \ z=0,d
\end{aligned}
\end{equation}
where $\textbf{n}$ is the unit outward normal to planes $z=0,d$.

Model (\ref{vad_mod}) admits the following steady solution $m_0$, which describes a situation with fluid at rest and heat spreading by conduction,
\begin{equation}\label{vad_eq1}
m_0\!=\!\left\{\textbf{v}_b \!=\! \textbf{0}\,, \,\,\,\, \bar{T_s} \!=\! \bar{T_f} \!=\! -\beta z \!+\! T_L\,, \quad p_b \!=\! -\rho_f g \alpha \beta \frac{z^2}{2}\! +\! \rho_f g \alpha T_L z \!+\! cost\right\}
\end{equation}
where $\beta=\dfrac{T_L-T_U}{d}(>0)$ is the adverse temperature gradient.

This paper is intended to investigate the stability of the $m_0$ solution with respect to perturbations to initial data. Therefore, we introduce the following perturbation fields $\{\textbf{u}, \theta, \phi, \pi\}$ on seepage velocity, fluid and solid temperature and pressure, respectively. The new solution of (\ref{vad_mod}) will be
\begin{equation}
\textbf{v} = \textbf{u} + \textbf{v}_b \qquad  T_f = \theta + \bar{T_f} \qquad T_s = \phi + \bar{T_s} \qquad p = \pi + p_b.
\end{equation}

Let us introduce the dimensionless quantities 
\begin{equation}
x_i =  x_i^* d, \quad t =  t^* \tau , \quad \pi = \pi^* P, \quad u_i = u^*_i U, \quad \theta = \theta^* T^{\prime}, \quad \phi = \phi^* T^{\prime}
\end{equation}
where
\begin{equation}
\tau=\frac{\rho  d^2c_\alpha}{\mu}\,,\quad P = \frac{U \mu d}{K_z}\,, \qquad U = \frac{\varepsilon \kappa_z^f}{(\rho c)_f d}\,, \qquad T^{\prime} = \beta d \sqrt{\frac{\kappa_z^f \varepsilon \mu}{\beta g \alpha K_z \rho_f^2 c_f d^2}},
\end{equation}
then, the dimensionless system for perturbation fields, omitting all the asterisks, is
\begin{equation}\label{vad_modad}
\begin{cases}
\mathcal{K}^{-1} \cdot \textbf{u} +\V^{-1} \textbf{u}_{,t}= -\nabla \pi + R \theta \textbf{k} - \mathcal{T} \textbf{k} \times \textbf{u} + Da \Delta \textbf{u}\\
\nabla \cdot \textbf{u}=0\\
\theta_{,t} + \textbf{u} \cdot \nabla \theta = R w + \eta \Delta_1 \theta + \theta_{,zz} + H(\phi - \theta)\\
A \phi_{,t} - \zeta \Delta_1 \phi - \phi_{,zz} + H \gamma (\phi - \theta) = 0
\end{cases}
\end{equation}
where 
\begin{equation*}
\gamma=\dfrac{\varepsilon \kappa_z^f}{(1-\varepsilon) \kappa_z^s}\,, \qquad A= \dfrac{(\rho c)_s \kappa_z^f}{(\rho c)_f \kappa_z^s}\,,\qquad  H=\dfrac{h d^2}{\varepsilon \kappa_z^f}
\end{equation*}
\begin{equation*}
\begin{aligned}
& R^2 = \dfrac{K_z \rho^2_f c_f d^2 \beta g \alpha}{\mu \varepsilon \kappa_z^f} \quad \mbox{Rayleigh number}, \qquad \mathrm{Da} = \frac{K_z \tilde{\mu}}{\mu d^2} \quad \mbox{Darcy number},\\
&  \qquad {\cal T}=\dfrac{2 \Omega \rho_f K_z}{\varepsilon \mu}\quad \mbox{Taylor number}, \qquad \V=\dfrac{c_f d^2 \mu}{K_z \kappa_z^f c_a} \quad \mbox{Vadasz number}.
\end{aligned}
\end{equation*}
System (\ref{vad_modad}) is coupled to the following initial conditions
\begin{equation}\label{vad_IC}
\textbf{u}(\textbf{x},0) \! = \!\textbf{u}_0(\textbf{x}) \,,\quad \theta(\textbf{x},0)\!=\!\theta_0(\textbf{x})\,,\quad \phi(\textbf{x},0)\!=\!\phi_0(\textbf{x})\,,\quad \pi(\textbf{x},0)\!=\!\pi_0(\textbf{x})
\end{equation} 
where $\nabla\cdot\textbf{u}_0=0$, and the following stress-free boundary conditions
\begin{equation}\label{vad_BC}
u_{,z} = v_{,z} = w = \theta = \phi = 0 \qquad \text{on} \ z=0,1.
\end{equation}

Let us assume that perturbations are periodic in $x$ and $y$ directions with periods $\dfrac{2\pi}{a_x}$ and $\dfrac{2\pi}{a_y}$, respectively. Let
\begin{equation}
V = \bigg[ 0, \dfrac{2\pi}{a_x} \bigg] \times \bigg[ 0, \dfrac{2\pi}{a_y} \bigg] \times [0, 1]
\end{equation}
be the periodicity cell, it is assumed that perturbations belong to $W^{2,2} (V), \, \forall t \in \mathbb{R}^+$ and they can be expanded as a Fourier series uniformly convergent in $V$.

\section{Linear instability}\label{vad_sec2}
In order to proceed to the linear instability analysis of the null solution of (\ref{vad_modad}), we take the curl and the double curl of $(\ref{vad_modad})_1$. By retaining only the third component of the resulting equations and by virtue of $(\ref{vad_modad})_2$, defining $\omega = \bm{\omega} \cdot \textbf{k}$ with $\bm{\omega} = \nabla \times \textbf{u}$ vorticity, we get
\begin{equation}\label{vad_mod2}
\begin{cases}
\xi^{-1} \omega + \V^{-1}  \omega_{,t} = \mathcal{T} w_{,z} + \mathrm{Da} \Delta \omega\\
\left( \xi \Delta_1 + \xi \V^{-1} \partial_{,t} \Delta_1 + \partial_{,zz} + \xi \V^{-1} \partial_{,t} \partial_{,zz} - \xi \mathrm{Da} \Delta \Delta \right) w =\\
 \qquad \qquad \qquad \qquad \qquad \qquad = \xi R \Delta_1 \theta - \xi \mathcal{T} \omega_{,z}
\end{cases}
\end{equation}
where $(\ref{vad_mod2})_2$ has been multiplied by $\xi$. By applying a further derivation with respect to $z$ and by multiplying by $\xi$, equation $(\ref{vad_mod2})_1$ becomes
\begin{equation}\label{vad_eq2}
\left( 1 + \xi \V^{-1} \partial_{,t} - \xi \mathrm{Da} \Delta \right) \omega_{,z} = \xi \mathcal{T} w_{,zz}
\end{equation}
As a consequence, we can apply the operator $\left( 1 + \xi \V^{-1} \partial_{,t} - \xi \mathrm{Da} \Delta \right)$ to $(\ref{vad_mod2})_2$ and plug (\ref{vad_eq2}) into the resulting equation so as to obtain
\begin{equation}
\begin{aligned}
& \left(\xi \Delta_1 + \partial_{,zz} + \xi \V^{-1} \partial_{,t} \Delta - \xi \mathrm{Da} \Delta \Delta \right) \left( 1 + \xi \V^{-1} \partial_{,t} - \xi \mathrm{Da} \Delta \right) w =\\
&\qquad \qquad \qquad \qquad = \xi R \Delta_1 \left( 1 + \xi \V^{-1} \partial_{,t} - \xi \mathrm{Da} \Delta \right) \theta - \xi^2 \mathcal{T}^2 w_{,zz}
\end{aligned}
\end{equation}
Hence, the linear version of model (\ref{vad_modad}) becomes
\begin{equation}\label{vad_modad1}
\begin{cases}
\begin{aligned}
& \left(\xi \Delta_1 + \partial_{,zz} + \xi \V^{-1} \partial_{,t} \Delta - \xi \mathrm{Da} \Delta \Delta \right) \left( 1 + \xi \V^{-1} \partial_{,t} - \xi \mathrm{Da} \Delta \right) w =\\
&\qquad \qquad \qquad \qquad = \xi R \Delta_1 \left( 1 + \xi \V^{-1} \partial_{,t} - \xi \mathrm{Da} \Delta \right) \theta - \xi^2 \mathcal{T}^2 w_{,zz}
\end{aligned}\\
\theta_{,t} = R w + \eta \Delta_1 \theta + \theta_{,zz} + H(\phi - \theta)\\
A \phi_{,t} - \zeta \Delta_1 \phi - \phi_{,zz} + H \gamma (\phi - \theta) = 0
\end{cases}
\end{equation}

System (\ref{vad_modad1}) is autonomous, then  solutions are such that the time dependence is separated from the  spatial one, i.e. 
\begin{equation}\label{vad_eq6}
\hat{\varphi} (t,\textbf{x}) = \varphi (\textbf{x}) \ e^{\sigma t} \quad \forall \hat{\varphi} \in \{w, \theta, \phi\} \qquad \sigma\in\C
\end{equation}
In addition, because of periodicity of perturbation fields, accounting for boundary conditions (\ref{vad_BC}) and since the sequence $\{ \sin n\pi z\}_{n\in\N}$ is a complete orthogonal system for $L^2([0,1])$, we can look for solution of (\ref{vad_modad1}) such that 
\begin{equation}\label{vad_eq8}
\varphi(x,y,z) = \sum_{n=1}^{+\infty} \bar{\varphi_n}(x,y,z)   \qquad \forall \varphi \in \{w, \theta, \phi \}
\end{equation}
where $\bar{\varphi_n} = \tilde{\varphi_n}(x,y) \sin(n \pi z)$ and  
\begin{equation}\label{vad_eq9}
\Delta_1 \bar{\varphi_n} = - a^2 \bar{\varphi_n}  \qquad  \partial_{,zz} \bar{\varphi_n}= - n^2 \pi^2 \bar{\varphi_n} \qquad (a^2=a^2_x + a^2_y)
\end{equation}
where $a$ is the wavenumber arising from spatial periodicity. 

\noindent
Now, let us define the following operators
\begin{equation}
\begin{aligned}
& \mathcal{L} \equiv \left( \xi \Delta_1 + \partial_{,zz} + \xi \V^{-1} \partial_{,t} \Delta - \xi \mathrm{Da} \Delta \Delta \right) \left( 1 + \xi \V^{-1} \partial_{,t} - \xi \mathrm{Da} \Delta\right) + \xi^2 \mathcal{T}^2 \partial_{,zz}\\
& \mathcal{L}_1 \equiv \partial_{,t} - \eta \Delta_1 - \partial_{,zz} + H\\
& \mathcal{L}_2 \equiv A \partial_{,t} - \zeta \Delta_1 - \partial_{,zz} + H \gamma 
\end{aligned}
\end{equation}
so that we can write (\ref{vad_modad1}) in the following way
\begin{equation}\label{vad_modad2}
\begin{cases}
\mathcal{L} w = \xi R \Delta_1 \left( 1 + \xi \V^{-1} \partial_{,t} - \xi \mathrm{Da} \Delta \right) \theta\\
\mathcal{L}_1 \theta = R w + H \phi \\
\mathcal{L}_2 \phi = H \gamma \theta
\end{cases}
\end{equation}
In order to get a single equation in the unknown $\theta$, let us apply the operator $\mathcal{L}_1$ and $\mathcal{L}_2$ to $(\ref{vad_modad2})_1$. Thus, one obtains
\begin{equation}
\mathcal{L} \mathcal{L}_1 \mathcal{L}_2 \theta = R \mathcal{L}_2 \mathcal{L} w + H \mathcal{L} \mathcal{L}_2 \phi
\end{equation}
which becomes
\begin{equation}\label{vad_eq3}
\mathcal{L} \mathcal{L}_1 \mathcal{L}_2 \theta =   \xi R^2 \Delta_1 \left( 1 + \xi \V^{-1} \partial_{,t} - \xi \mathrm{Da} \Delta \right)\mathcal{L}_2 \theta + H^2 \gamma \mathcal{L} \theta
\end{equation}
by virtue of $(\ref{vad_modad2})_2$-$(\ref{vad_modad2})_3$. Then, we split $\mathcal{L}_2$ in the first term of (\ref{vad_eq3})
\begin{equation}\label{vad_eq4}
\begin{aligned}
& \mathcal{L} \mathcal{L}_1 \left( A \partial_{,t} - \zeta \Delta_1 - \partial_{,zz} \right) \theta + H \gamma \mathcal{L} \left( \partial_{,t} - \eta \Delta_1 - \partial_{,zz} \right)\theta = \\
& \qquad \qquad \qquad \qquad \qquad =  \xi R^2 \Delta_1 \left( 1 + \xi \V^{-1} \partial_{,t} - \xi \mathrm{Da} \Delta \right)\mathcal{L}_2 \theta
\end{aligned}
\end{equation}
and we split $\mathcal{L}_1$ in the first term of (\ref{vad_eq4}) 
\begin{equation}\label{vad_eq5}
\begin{aligned}
& \mathcal{L} \left( \partial_{,t} - \eta \Delta_1 - \partial_{,zz} \right)\mathcal{L}_2 \theta + H \mathcal{L} \left( A \partial_{,t} - \zeta \Delta_1 - \partial_{,zz} \right) \theta  = \\
& \qquad\qquad\qquad\qquad\qquad  \xi R^2 \Delta_1 \left( 1 + \xi \V^{-1} \partial_{,t} - \xi \mathrm{Da} \Delta \right)\mathcal{L}_2 \theta
\end{aligned}
\end{equation}
From (\ref{vad_eq5}), we can determine the critical value for the Rayleigh number beyond which either steady or oscillatory convection occurs. To this aim, we substitute (\ref{vad_eq6})-(\ref{vad_eq8}) in (\ref{vad_eq5}) and retain only the $n$-th component. As a consequence, we obtain
\begin{equation}\label{vad_eq7}
\begin{aligned}
& [-\xi a^2 - n^2 \pi^2 - \xi\mathrm{Da} \delta_n^2 - \xi^2 \mathrm{Da} \delta_n a^2 - \xi \mathrm{Da} n^2 \pi^2\delta_n - \xi^2 \mathrm{Da}^2 \delta_n^3 - \xi^2 \mathcal{T}^2 n^2 \pi^2 +\\
& + \sigma \xi \V^{-1} \left( -\delta_n -\xi a^2 -n^2 \pi^2 - 2 \xi \mathrm{Da}\delta_n^2 \right) -\sigma^2 \xi^2 \V^{-2}\delta_n ]\cdot\\
&  \left[ \left( \sigma + \eta a^2 +n^2\pi^2 \right)\left( A \sigma +\zeta a^2 +n^2 \pi^2+H\gamma \right) + H \left( A \sigma +\zeta a^2 +n^2\pi^2 \right)\right] =\\
& = -\xi R^2 a^2 \left( A \sigma +\zeta a^2 +n^2\pi^2+H\gamma \right) \left( 1 + \sigma \xi \V^{-1} + \xi \mathrm{Da} \delta_n \right) 
\end{aligned}
\end{equation}
being $\delta_n=a^2 + n^2 \pi^2$.

Accounting for \eqref{vad_eq7}, one can simply deduce that the eigenvalue $ \sigma $ can assume pure imaginary values, meaning that the principle of exchange of stabilities does not hold. The illuminating paper by Vadasz \cite{Vadasz1998} suggests that this is due to the  presence of the Vadasz inertia term in the momentum equation $(\ref{vad_modad})_1$.  For this reason, we can claim that the Vadasz number (coupled to the action of rotation) allows convection to arise either via oscillatory or steady motions .

It is well known that instability occurs once the eigenvalue crosses either the axis of pure imaginary numbers or the zero value. Therefore, we set once $\sigma =0$ and once $\sigma =i\omega \ (\omega \in \R-\{0\})$ in (\ref{vad_eq7}).

\subsection{Steady convection}
In order to determine the critical Rayleigh number for steady convection, we set $\sigma =0$ in (\ref{vad_eq7}) so as to obtain
\begin{equation}\label{vad_eq10}
\begin{aligned}
& R^2 = F(n^2,a^2) =\\
& = \frac{ \left( 1 + \xi \mathrm{Da} \delta_n \right) \left( \xi a^2 + n^2 \pi^2 + \xi\mathrm{Da} \delta_n^2 \right) + \xi^2 \mathcal{T}^2 n^2 \pi^2 }{\xi a^2  \left( 1 + \xi \mathrm{Da} \delta_n \right)}\\
&  \left[ \eta a^2 +n^2\pi^2 + H \frac{\zeta a^2 +n^2 \pi^2}{\zeta a^2 +n^2 \pi^2+H\gamma}\right] 
\end{aligned}
\end{equation}
Starting from (\ref{vad_eq10}), the critical threshold $R_S$ is determined by solving the minimum problem
\begin{equation}
R_S = \min_{(n^2,a^2) \in \N\times\R^+} F(n^2,a^2)
\end{equation}
Since $F(n^2,a^2)$ is a strictly increasing function with respect to $n^2$, the critical threshold is
\begin{equation}\label{vad_eq13}
R_S = \min_{a^2 \in \R^+} F(1,a^2)
\end{equation}
Let us remark that
\begin{itemize}
\item $R_S$ does not depend on the Vadasz number $\V$. This means that steady convection is not affected by inertia effects. Indeed, the critical threshold $R_S$ coincides with the steady threshold found by \cite{Gianfrani2021b};
\item if the porous medium is isotropic, i.e. $\xi=\eta=\zeta=1$, the critical threshold $R_S$ is the same as that one found by \cite{Malashetty2010};
\item if the medium porosity is low, namely the Brinkmann model is no longer suitable to describe the fluid motion (i.e. $\mathrm{Da}=0$), $R_S$ coincides with the steady threshold determined by 
\cite{Shiv2015}, where the authors studied the Darcy model. In addition, if the medium is isotropic, $R_S$ reduces to the threshold found by \cite{Mal2007};
\item it is straightforward to notice that $R_S$ is a strictly increasing function with respect to $\mathcal{T}$ and $\eta$, which implies that rotation and fluid thermal conductivity have a delaying effect on the onset of convection. Physical meaning of this behaviour is pointed out in Section \ref{vad_sec3};
\item since the derivative of $F(1, a^2)$ with respect to $\xi$ is
\begin{equation}
\partial_{\xi} F(1,a^2) = \xi^2 \left( \mathcal{T}^2 - \mathrm{Da}^2 \delta^2 \right) - 2 \xi \mathrm{Da} \delta - 1
\end{equation}
being $\delta=a^2+\pi^2$, the behaviour of $R_S$ with respect to $\xi$ depends on $\mathcal{T}$. In particular, if $\mathcal{T}=0$, the derivative is negative. Such a result proves analytically the stabilising effect of permeability on the onset of convection in absence of rotation.
\end{itemize}

\subsection{Oscillatory convection}
In order to determine the critical Rayleigh number for oscillatory convection, we set $\sigma =i\omega \ (\omega \in \R-\{0\})$ in (\ref{vad_eq7}) 
\begin{equation}\label{vad_eq11}
\begin{aligned}
& R^2 = G(n^2,a^2)=\\
& = \frac{\left( 1 + i\omega \xi \V^{-1} + \xi \mathrm{Da} \delta_n \right)  \left( \xi a^2 + n^2 \pi^2 + \xi\mathrm{Da} \delta_n^2 + i \omega \xi \V^{-1} \delta_n \right) + \xi^2 \mathcal{T}^2 n^2 \pi^2 }{\xi a^2 \left( 1 + i\omega \xi \V^{-1} + \xi \mathrm{Da} \delta_n \right)}\\
&  \left[ i\omega + \eta a^2 +n^2\pi^2 + H \frac{ A i\omega +\zeta a^2 +n^2 \pi^2}{ A i\omega +\zeta a^2 +n^2 \pi^2+H\gamma} \right]
\end{aligned}
\end{equation}
The Rayleigh number is a real number, that is why the imaginary part in (\ref{vad_eq11}) has to be set equal to zero. By doing so, set $\omega_*=\omega^2$, the following equation arises 
\begin{equation}\label{vad_eq12}
\frac{J_1 \omega_*^2 + J_2 \omega_* + J_3}{ a^2 \left[\left(\delta_{\zeta} + \gamma H\right)^2 + 
      A^2 \omega_*\right] \V \xi \left[\omega_* \xi^2 + \left(\V + \mathrm{Da} \delta_n \V \xi\right)^2\right]} =0
\end{equation}
which provides a condition for the existence of oscillatory convection, where
\begin{equation}\label{vad_eq12b}
\begin{aligned}
J_1 = & A^2 \xi^2 \left[ H \xi \delta_n + \V  \left( \xi a^2 + n^2 \pi^2 + \delta_n^2 \xi \mathrm{Da} \right) + \xi \delta_n \left( \eta a^2 + n^2 \pi^2 \right) \right]\\
J_2 = & A^2 \V^2 \{ \xi (\mathrm{Da} \delta_n \xi+1)^2 \left[a^2 \V+\delta_n (\mathrm{Da} \delta_n \V+\delta_{\eta}+H)\right]+ \\
& \pi^2 n^2 \left[\V (\mathrm{Da} \delta_n \xi+1) \left(\mathrm{Da} \delta_n \xi+T^2 \xi^2+1\right)-T^2 \xi^3 (\delta_{\eta}+H)\right]\}+\\
& A \gamma H^2 \V \xi^2 \left[\xi \left(a^2+\mathrm{Da} \delta_n^2\right)+\pi ^2 n^2\right]+\xi^2 (\delta_{\zeta}+\gamma H)\\
& \{\xi \left[\V \left(a^2+\mathrm{Da} \delta_n^2\right) \left(\delta_{\zeta}+\gamma H\right)+\delta_n \delta_{\eta} (\delta_{\zeta}+\gamma H)+\delta_n \delta_{\zeta} H\right]+\\
& \pi ^2 n^2 \V \left(\delta_{\zeta}+\gamma H\right)\}\\
J_3 = &  \V^2 \gamma H^2 \{ A \V (\mathrm{Da} \delta_n \xi+1) [\xi \left(a^2+\mathrm{Da} \delta_n^2\right) \left(\mathrm{Da} \delta_n \xi+1\right)+\\
& \pi^2 n^2 \left(\mathrm{Da} \delta_n \xi+T^2 \xi^2+1\right)]+\\
& \gamma [ \xi (\mathrm{Da} \delta_n \xi+1)^2 \left(a^2 \V+\delta_n (\mathrm{Da} \delta_n \V+\delta_{\eta})\right)+\\
& \pi^2 n^2 \left(\V (\mathrm{Da} \delta_n \xi+1) \left(\mathrm{Da} \delta_n \xi+T^2 \xi^2+1\right)-\delta_{\eta} T^2 \xi^3\right)]\}+\\
& \V^2 \delta_{\zeta}^2 \{ \xi (\mathrm{Da} \delta_n \xi+1)^2 \left[a^2 \V+\delta_n (\mathrm{Da} \delta_n \V+\delta_{\eta}+H)\right]+\\
& \pi ^2 n^2 \left[\V (\mathrm{Da} \delta_n \xi+1) \left(\mathrm{Da} \delta_n \xi+T^2 \xi^2+1\right)-T^2 \xi^3 (\delta_{\eta}+H)\right]\}+\\
& \V^2\delta_{\zeta} \gamma H \{ \xi (\mathrm{Da} \delta_n \xi+1)^2 \left[2 a^2 \V+\delta_n (2 \mathrm{Da} \delta_n \V+2 \delta_{\eta}+H)\right]+\\
& \pi ^2 n^2 \left[2 \V (\mathrm{Da} \delta_n \xi+1) \left(\mathrm{Da} \delta_n \xi+T^2 \xi^2+1\right)-T^2 \xi^3 (2 \delta_{\eta}+H)\right]\}
\end{aligned}
\end{equation}
and 
\begin{equation}
\begin{aligned}
& \delta_{\eta} = \eta a^2 +n^2 \pi^2\\
& \delta_{\zeta} = \zeta a^2 +n^2 \pi^2
\end{aligned}
\end{equation}
Let us notice that, since $J_1$ is always positive, oscillatory convection cannot occur when either
\begin{equation}\label{vad_eq17}
J_2^2-4 J_1 J_3 <0
\end{equation}
or 
\begin{equation}\label{vad_eq18}
\begin{cases}
J_2>0\\
J_3>0
\end{cases}
\end{equation}
Once the positive root of (\ref{vad_eq12}) has been determined, it is substituted in (\ref{vad_eq11}) and the critical Rayleigh number for the onset of oscillatory convection is determined by solving the following minimum problem:
\begin{equation}\label{vad_eq19}
R_O = \min_{(n^2,a^2) \in \N\times\R^+} G_1(n^2,a^2, \omega^2(n^2,a^2))|_{\omega^2=\omega^2_c}
\end{equation}   
 $\omega_c^2$ being such that $Im(G(n^2,a^2,\omega^2_c(n^2,a^2)))=0$, and
\begin{equation}\label{vad_eq19b}
\begin{aligned}
& G_1\left(n^2,a^2,\omega^2(n^2,a^2)\right) =  Re \left(G\left(n^2,a^2,\omega^2(n^2,a^2)\right)\right)=\\
& \left[\delta_{\zeta} H + \delta_{\eta} (\delta_{\zeta} + \gamma H) - A \omega^2\right] \left[ -A \omega^2 \xi + (\delta_{\zeta} + \gamma H) \V (1 + \mathrm{Da} \delta_n \xi)\right]\\
& \left[-\delta_n \omega^2 \xi^2 + n^2 \pi^2 \mathcal{T}^2 \V^2 \xi^2 + \V^2 (1 +  \mathrm{Da} \delta_n \xi) \left(n^2 \pi^2 + \xi a^2 + \mathrm{Da} \delta_n^2 \xi\right)\right] + \\
&  \bigg[\left[\delta_{\zeta} H + 
         \delta_{\eta} (\delta_{\zeta} + \gamma H) - A \omega^2\right] \V \xi \left(\delta_n + 
         n^2 \pi^2 + a^2 \xi + 2 \mathrm{Da} \delta_n^2 \xi\right) \\
& \left[A \V + (\delta_{\zeta} + \gamma H + A \mathrm{Da} \delta_n \V) \xi\right] + \left[\delta_{\zeta} + \gamma H + A (\delta_{\eta} + H)\right] \V \xi \\
& \left(\delta_n + n^2 \pi^2 + a^2 \xi + 
         2 \mathrm{Da} \delta_n^2 \xi\right) \left[A \omega^2 \xi - (\delta_{\zeta} + \gamma H) \V (1 + \mathrm{Da} \delta_n \xi)\right] + \\
& \left[\delta_{\zeta} + \gamma H + 
         A (\delta_{\eta} + H)\right] \left[A \V + \xi(\delta_{\zeta} + \gamma H + 
            A \mathrm{Da} \delta_n \V) \right] \\
&\left[-\delta_n \omega^2 \xi^2 + 
         n^2 \pi^2 \mathcal{T}^2 \V^2 \xi^2 + \V^2 (1 + 
            \mathrm{Da} \delta_n \xi) (n^2 \pi^2 + \xi a^2 + 
               \mathrm{Da} \delta_n^2 \xi)\right]\bigg]\\
& \dfrac{\omega^2}{a^2 \left[\left(\delta_{\zeta} + \gamma H\right)^2 + A^2 \omega^2\right] \V \xi \left[\omega^2 \xi^2 + \left(\V + \mathrm{Da} \delta_n \V \xi\right)^2\right]}
\end{aligned}
\end{equation}

We would like to remark that if (\ref{vad_eq12}) admits two positive roots, then both of them are plugged into (\ref{vad_eq11}) and the lowest threshold arising from the two is the critical value we were looking for.

Accounting for the physical meaning of $ R_O $, we have to prove that $G_1\left(n^2,a^2, \omega^2(n^2,a^2)\right)$ is positive. First, we write $G_1$ as follows so that it is easier to understand where this functions lives:
\begin{equation}
G_1\left(n^2,a^2, \omega^2(n^2,a^2)\right) = \dfrac{\Lambda_1 \omega_*^3 + \Lambda_2 \omega_*^2 + \Lambda_3 \omega_* + \Lambda_4}{\Lambda_5 \omega_*^2+\Lambda_6 \omega_* + \Lambda_7}
\end{equation}
where
\begin{equation}
\begin{aligned}
\Lambda_1 = & -A^2 \delta \xi^3\\
\Lambda_2 = & \, a^2 A^2 \delta_{\eta} \V \xi^3+a^2 A^2 H \V \xi^3-A^2 \mathrm{Da}^2 \delta^3 \V^2 \xi^3+A^2 \mathrm{Da} \delta^2 \delta_{\eta} \V \xi^3+\\
& A^2 \mathrm{Da} \delta^2 H \V \xi^3-2 A^2 \mathrm{Da} \delta^2 \V^2 \xi^2-A^2 \delta \V^2 \xi+\pi^2 A^2 \delta_{\eta} n^2 \V \xi^2+\\
& \pi^2 A^2 H n^2 \V \xi^2+ \pi^2 A^2 n^2 \mathcal{T}^2 \V^2 \xi^3-A \delta \gamma H^2 \xi^3-\delta \delta_{\zeta}^2 \xi^3-\\
& 2 \delta \delta_{\zeta} \gamma H \xi^3-\delta \gamma^2 H^2 \xi^3\\
\Lambda_3 = & \, \V \bigg\{A^2 \V^2 (\delta_{\eta}+H) (\mathrm{Da} \delta \xi+1) \{\xi (a^2+\mathrm{Da} \delta^2) (\mathrm{Da} \delta \xi+1)+\\
& \pi^2 n^2 \left(\mathrm{Da} \delta \xi+\mathcal{T}^2 \xi^2+1\right)\}+\\
& \xi (\delta_{\zeta}+\gamma H) \bigg[\xi^2 \bigg( (a^2+\mathrm{Da} \delta^2) [\delta_{\eta} (\delta_{\zeta}+\gamma H)+\delta_{\zeta} H]-\\
& \V (\delta_{\zeta}+\gamma H) (\mathrm{Da}^2 \delta^3-\pi^2 n^2 \mathcal{T}^2)\bigg)-\delta \V (\delta_{\zeta}+\gamma H)+ \\
& \xi \bigg(\pi^2 n^2 [\delta_{\eta} (\delta_{\zeta}+\gamma H)+\delta_{\zeta} H]-2 \mathrm{Da} \delta^2 \V (\delta_{\zeta}+\gamma H)\bigg)\bigg]-\\
& A \gamma H^2 \V \xi \left[\xi^2 \left(\mathrm{Da}^2 \delta^3-\pi^2 n^2 \mathcal{T}^2\right)+2 \mathrm{Da} \delta^2 \xi+\delta\right]\bigg\}\\
\Lambda_4 = & \, \V^3 (\mathrm{Da} \delta \xi+1) (\delta_{\zeta}+\gamma H) \left[\delta_{\eta} (\delta_{\zeta}+\gamma H)+\delta_{\zeta} \right]\\
& \left[(\xi a^2+\mathrm{Da} \delta^2\xi) (\mathrm{Da} \delta \xi+1)+\pi^2 n^2 \left(\mathrm{Da} \delta \xi+\mathcal{T}^2 \xi^2+1\right)\right]\\
\Lambda_5 = & \, a^2 A^2  \V \xi^3\\
\Lambda_6 = & \, a^2 \V \xi \left[A^2 (\mathrm{Da} \delta \V \xi+\V)^2+\xi^2 (\delta_{\zeta}+\gamma H)^2\right]\\
\Lambda_7 = & \, a^2 \V^3 \xi (\mathrm{Da} \delta \xi+1)^2 (\delta_{\zeta}+\gamma H)^2
\end{aligned}
\end{equation}
It is straightforward to notice that the denominator is always positive, as well as $\Lambda_4$, while $\Lambda_1<0$. Despite we know nothing about the sign of $\Lambda_2$ and $\Lambda_3$, we can say that $G_1$ lives in the first quarter for any $\omega^2 \in [0,\bar{\omega}^2]$, where $\bar{\omega}^2$ such that $G_1(n^2,a^2,\bar{\omega}^2)=0$. Numerically, it is easy to notice that the following system
\begin{equation}
\begin{cases}
Im(G(n^2,a^2,\omega^2(n^2,a^2)))=0\\
G_1(n^2,a^2,\omega^2(n^2,a^2))=0
\end{cases}
\end{equation}
does not admit any solution $\omega^2(n^2,a^2)$. Furthermore, and most importantly, numerical simulations show that $\omega^2_c(n^2,a^2) < \bar{\omega}^2(n^2,a^2)$, $\forall (n^2, a^2)$. Hence, $G_1(n^2,a^2,\omega_c^2(n^2,a^2))>0$, $\forall (n^2, a^2)$.

\begin{remark}
	If $\mathcal{T}=0$, from (\ref{vad_eq12b}) it turns out that $J_2, J_3 >0$. Hence, by virtue of (\ref{vad_eq18}), conditions for the existence of oscillatory convection are not satisfied and only steady convection can occur. 
\end{remark}

\begin{remark}\label{vad_rem2}
	If $\V \rightarrow \infty$, by performing the limit in (\ref{vad_eq12}), it turns out that (\ref{vad_eq12}) admits only negative solution. As a consequence, only steady convection can occur and the analysis is the same as the one presented in \cite{Gianfrani2021b}.  
\end{remark}

\section{Numerical simulation} \label{vad_sec3}
Given the complex expression of the critical thresholds, both for oscillatory and steady convection, a numerical analysis is required in order to draw the attention on how parameters affect the onset of instability. In particular, in this section, we will point out the effect of rotation, the Darcy number and anisotropic permeability and thermal conductivities on the motionless steady state. A particular attention will be paid to the effect of Vadasz inertia term, which can cause the onset of convection via oscillatory motion, as already pointed out.

Before proceeding to the analysis, let us underline that, due to its complexity, it is not possible to solve analytically the minimum problem (\ref{vad_eq19})-(\ref{vad_eq19b}). Numerical simulations show that the minimum for $G_1(n^2,a^2, \omega_c^2(n^2,a^2))$ with respect to $n^2$ is attained in $n^2=1$, for a completely arbitrary choice of parameters. That is why we can reduce our analysis to the following minimum problem
\begin{equation}\label{vad_eq20}
R_O = \min_{a^2 \in \R^+} G_1(1,a^2, \omega_c^2(1,a^2))
\end{equation}
Once $R_O$ has been determined, we can define the critical Rayleigh number $Ra$ as
\begin{equation}
Ra= \min \{R_S, R_O \}
\end{equation}
If $Ra=R_S$ ($Ra=R_O$), convection can occur through steady (oscillatory) motions.

In order to show the influence of parameters on the critical Rayleigh number $Ra$, we adopt the same procedure as done in \cite{Gianfrani2020} - \cite{Gov}, namely we report the critical threshold as function of the inter-phase heat transfer coefficient $H$. This parameter, which is defined by $H=\dfrac{h d^2}{\varepsilon \kappa_z^f}$, is not always easily measured. That is why we need to introduce a range of values between which $H$ can vary. Starting from its definition, $(10^{-2}, 10^{6})$ is a reasonable interval.

\begin{figure}[h!]
	\subfloat[]{
		\includegraphics[scale=0.4]{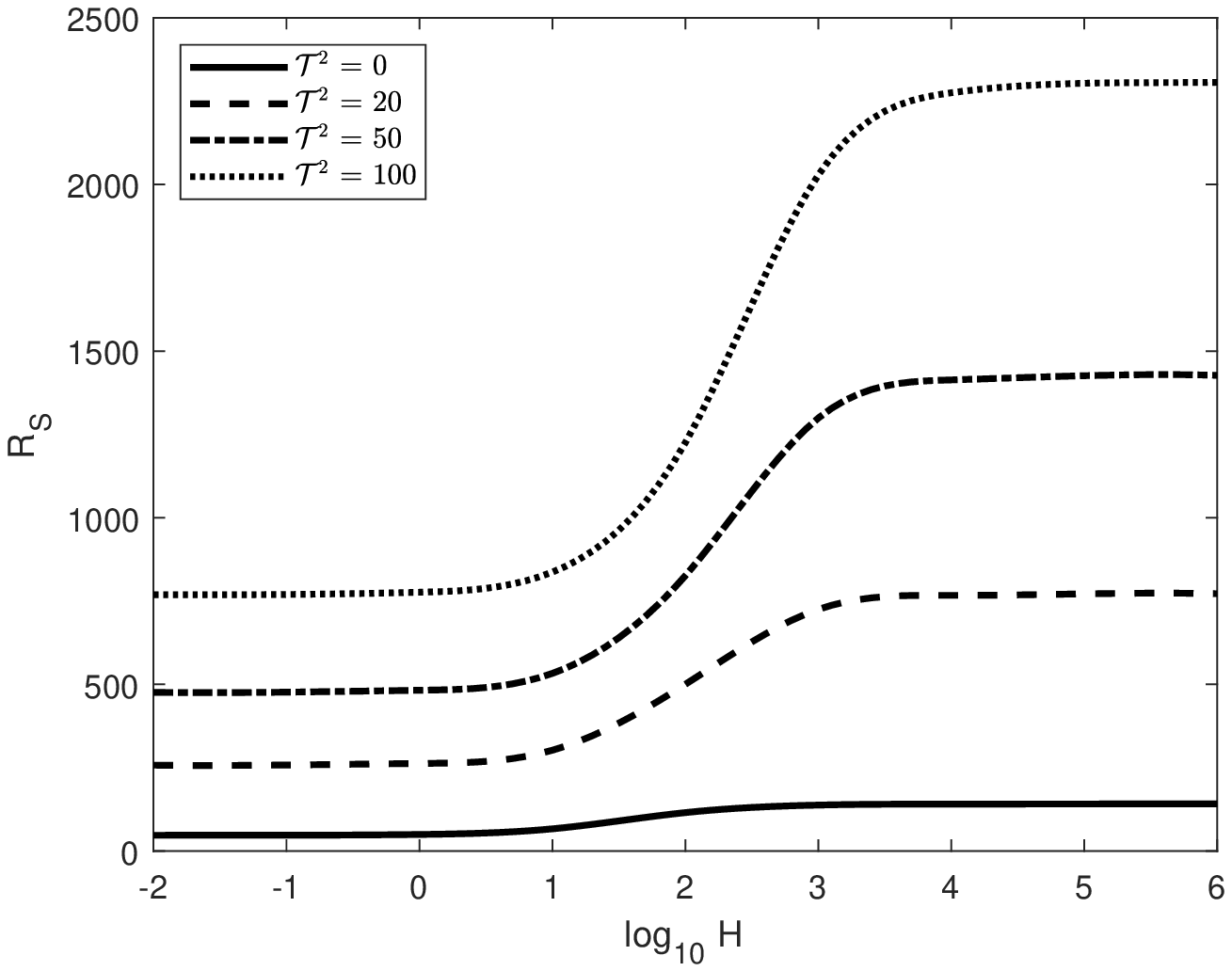}
		\label{vad_fig1a}}
	\subfloat[]{
		\includegraphics[scale=0.4]{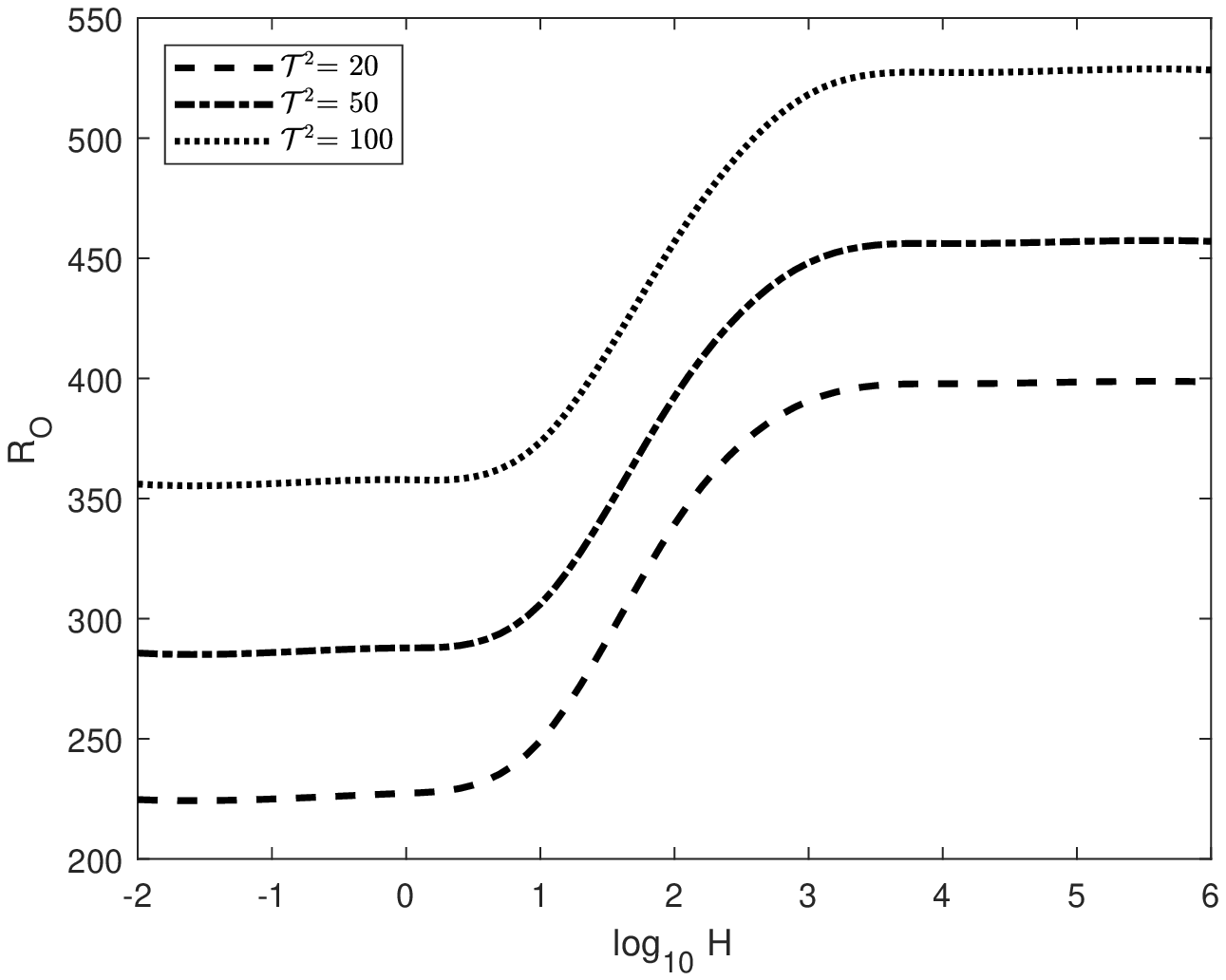}
		\label{vad_fig1b}}
	\caption{Critical Rayleigh number as function of the inter-phase heat transfer coefficient $H$ for different values of the Taylor number $\mathcal{T}$ with $\xi=\zeta = \eta =1$, $\gamma = 0.5$, $A=0.01$, $\V=10$ and $\mathrm{Da}=0.01$.}
	\label{vad_fig1}
\end{figure}

\begin{figure}[h!]
	\subfloat[]{
		\includegraphics[scale=0.4]{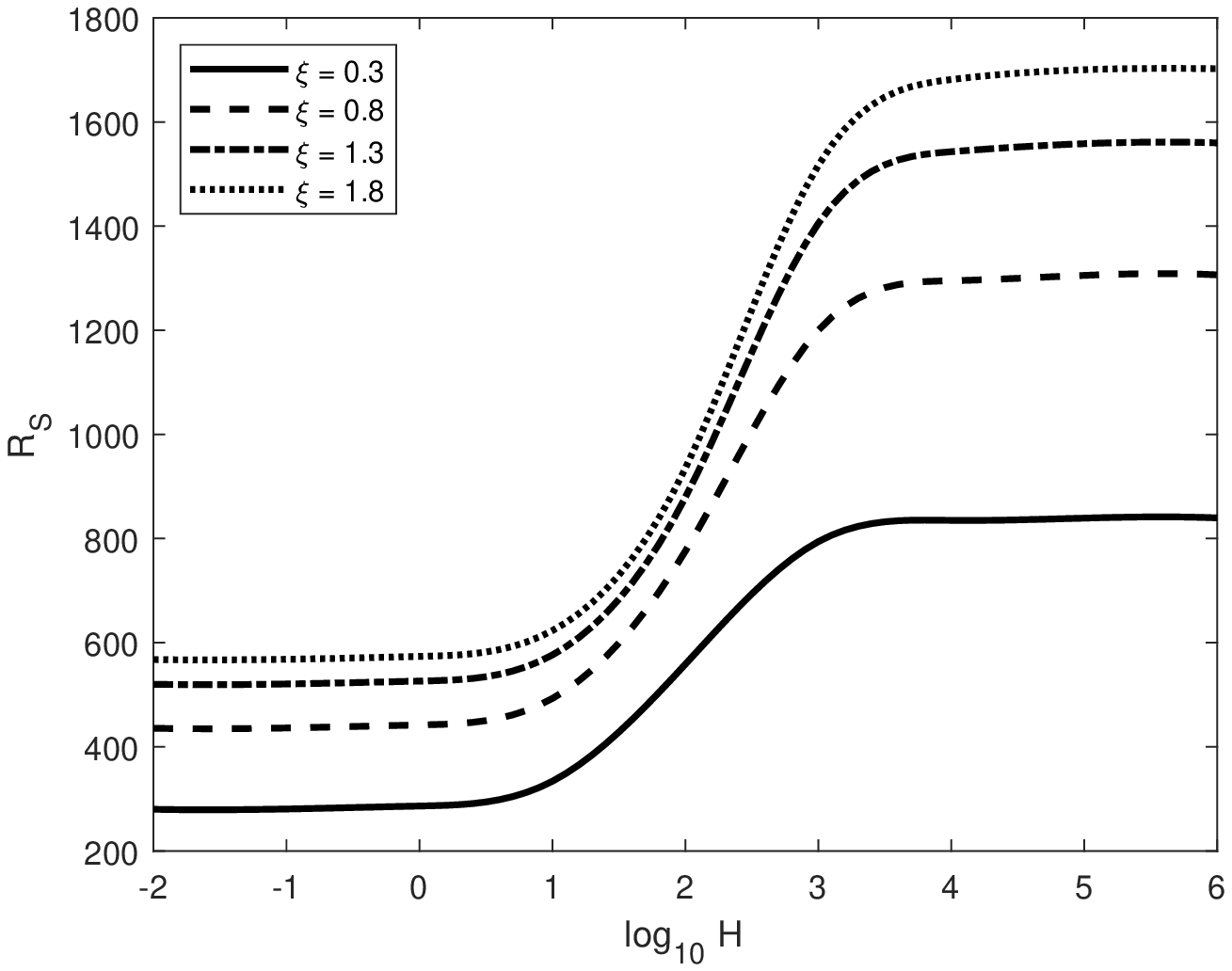}
		\label{vad_fig2a}}
	\subfloat[]{
		\includegraphics[scale=0.4]{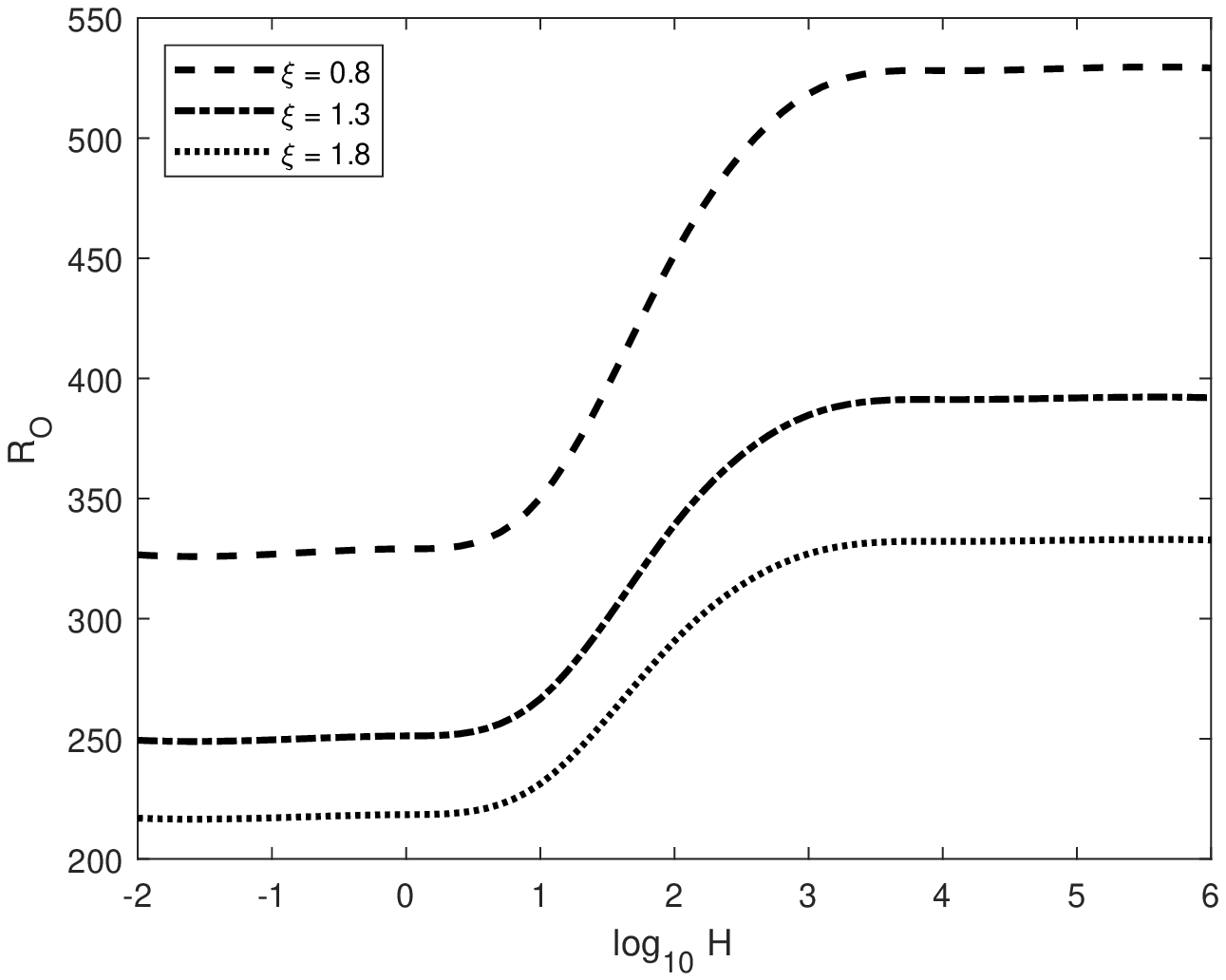}
		\label{vad_fig2b}}
	\caption{Critical Rayleigh number as function of the inter-phase heat transfer coefficient $H$ for different values of permeability $\xi$ with $\zeta = \eta =1$, $\gamma = 0.5$, $A=0.01$, $\V=10$, $\mathcal{T}^2=50$ and $\mathrm{Da}=0.01$.}
	\label{vad_fig2}
\end{figure}

Figures \ref{vad_fig1a}-\ref{vad_fig1b} show the influence of rotation on $R_S$ and $R_O$, respectively. The delaying effect of rotation on the onset of convection comes out. We have previously shown analytically that the critical threshold $R_S$ for steady convection increases with $\mathcal{T}$, so the result in Figure \ref{vad_fig1a} is expected. In addition, what Figures \ref{vad_fig1a}-\ref{vad_fig1b} show is physically reasonable because in the momentum equation $(\ref{vad_modad})_1$ the term due to rotation has only horizontal component, namely rotation acts horizontally on the fluid motion, discouraging the motion in the vertical direction. 
We managed to prove that, when $\mathcal{T}=0$, conditions for the existence of oscillatory convection are not satisfied. As a consequence, in Figure \ref{vad_fig1b}, if $\mathcal{T}=0$, the plot of $R_O$ does not appear. This result is in agreement with results in \cite{Shiv2011} and, in addition, it is expected since the inertia term leads to the occurrence of oscillatory convection only when coupled to the rotation term, as already pointed out \cite{Vadasz1998}.

\begin{figure}[h!]
	\includegraphics[scale=0.4]{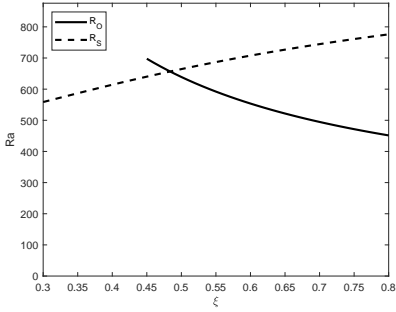}
	\caption{Critical Rayleigh number as function of the permeability parameter $\xi$ with $\zeta = \eta =1$, $\gamma = 0.5$, $H=100$, $A=0.01$, $\V=10$, $\mathcal{T}^2=50$ and $\mathrm{Da}=0.01$.}
	\label{vad_fig3}
\end{figure}

In Figures \ref{vad_fig2a}-\ref{vad_fig2b} the behaviour of $R_S$ and $R_O$ with respect to permeability $\xi$ is shown. It is evident the different influence of $\xi$ on the onset of oscillatory and steady convection. Increasing the horizontal permeability makes the onset of oscillatory motions easier, discouraging the onset of steady ones. Actually, since we managed to determine the derivative $\partial_{\xi} F(1, a^2)$, we proved analytically that the behaviour of $R_S$ with respect to $\xi$ depends on $\mathcal{T}$. The destabilizing effect of $\xi$ on the onset of steady convection has been proved analytically if $\mathcal{T}=0$. While if $\mathcal{T}\neq 0$, $\xi$ keeps on encouraging the onset of instability, but, unlike the previous case, oscillatory motions are preferred instead of steady ones, as shown in Figure \ref{vad_fig3}.
In Table \ref{vad_tab1}, the behaviour of the critical Rayleigh number $Ra$ with respect to $\xi$ is shown, both with and without rotation. In particular, in Table \ref{vad_tab1a}, the values of $Ra$ are bold. While in Table \ref{vad_tab1b}, given the absence of oscillatory convection, $Ra$ coincides with $R_S$. 


\begin{table}[h!]
	\centering
	\subfloat[$\mathcal{T}^2=50$]{
		\begin{tabular}{lll}
			\hline\noalign{\smallskip}
			$\xi$ & $R_O$ & $R_S$ \\
			\noalign{\smallskip}\hline\noalign{\smallskip}
			0.30	&	$\nexists$	&	\textbf{558.6386}\\
			0.42	&	$\nexists$	&	\textbf{624.7509}\\
			0.43	&	725.4627	&	\textbf{629.9432}\\
			0.48	&	660.8735	&	\textbf{654.8910}\\
			0.49	&	\textbf{649.7135}	&	659.6715\\
			0.70	&	\textbf{494.7303}	&	744.5317\\
			\noalign{\smallskip}\hline
		\end{tabular} \label{vad_tab1a} 
	}
	\qquad
	\subfloat[$\mathcal{T}^2=0$]{
		\begin{tabular}{ll}
			\hline\noalign{\smallskip}
			$\xi$ & $R_S \equiv Ra$ \\
			\noalign{\smallskip}\hline\noalign{\smallskip}
			0.30	&	207.4724\\
			0.35	&	190.6250\\
			0.40	&	177.5048\\
			0.50	&	158.2614\\
			0.60	&	144.7119\\
			0.70	&	134.5796\\
			\noalign{\smallskip}\hline
		\end{tabular} \label{vad_tab1b} 
	}
	\caption{Critical Rayleigh number $Ra$ for different values of $\xi$ with $\zeta=\eta=1$, $A=0.01$, $\gamma=0.5$, $H=100$, $\mathrm{Da} =0.01$, $\V=10$, (a) $\mathcal{T}^2=50$ (b) $\mathcal{T}=0$.}
	\label{vad_tab1}
\end{table}

\begin{figure}[h!]
	\subfloat[]{
		\includegraphics[scale=0.4]{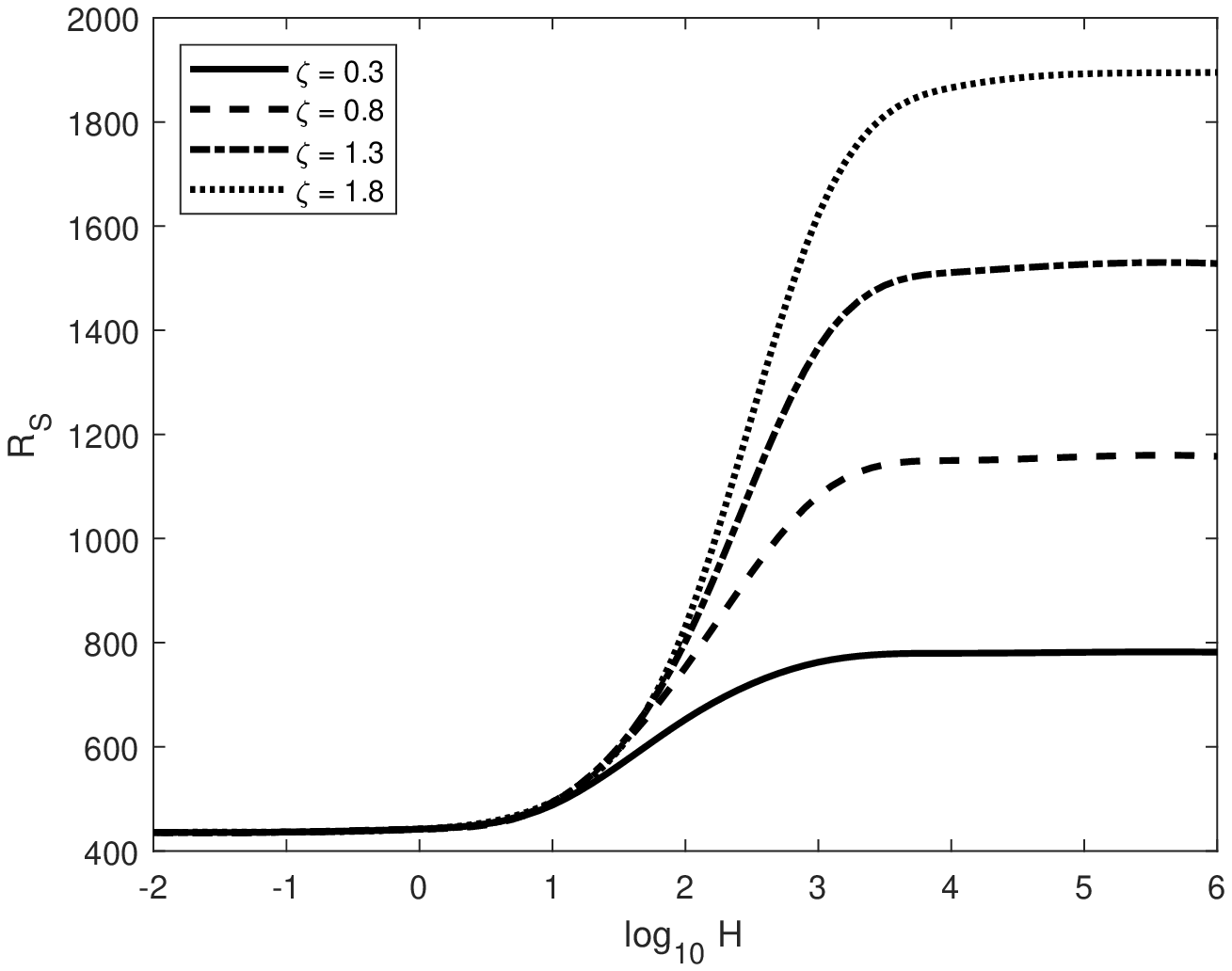}
		\label{vad_fig4a}}
	\subfloat[]{
		\includegraphics[scale=0.4]{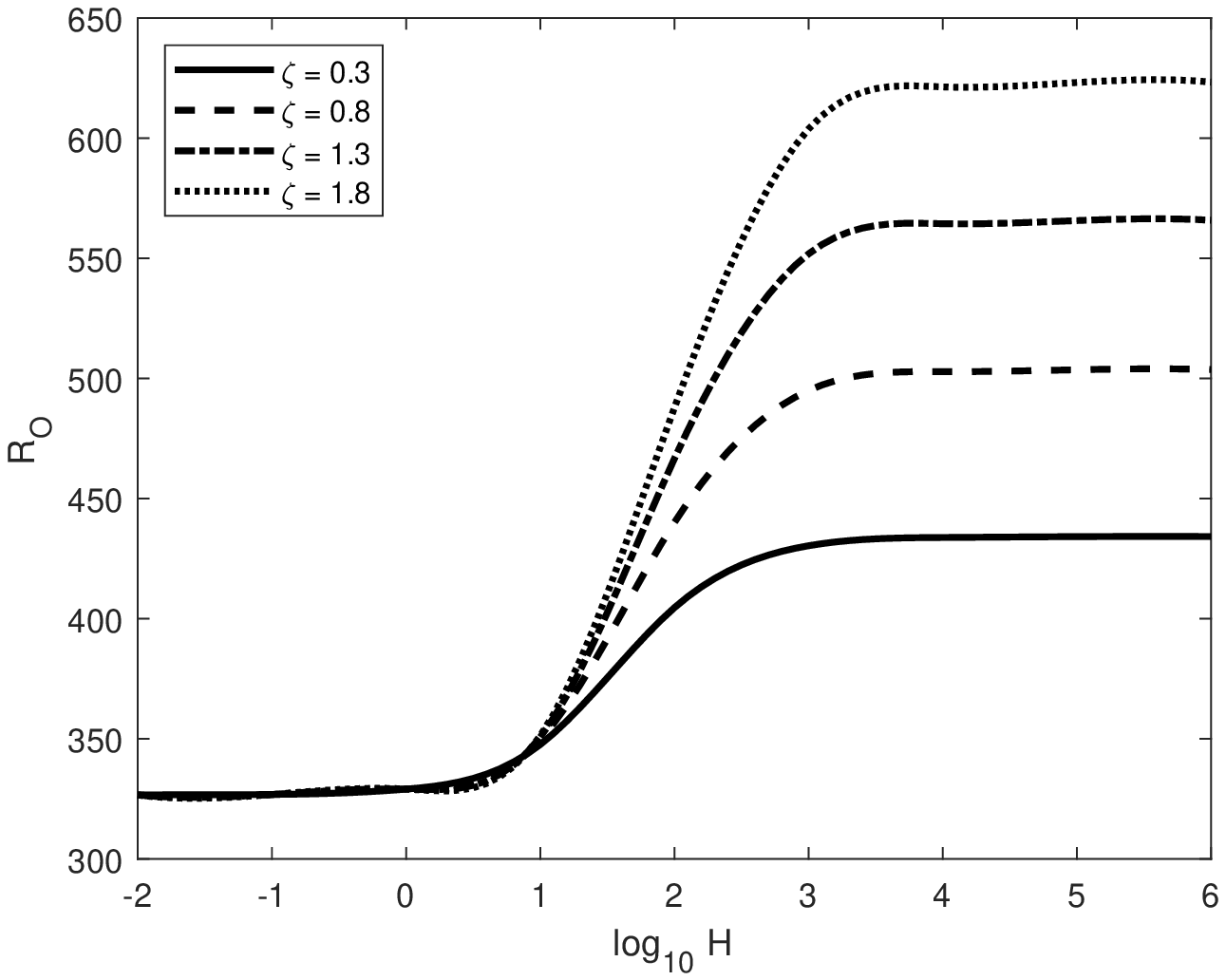}
		\label{vad_fig4b}}
	\caption{Critical Rayleigh number as function of the inter-phase heat transfer coefficient $H$ for different values of the solid thermal conductivity parameter $\zeta$ with $\xi = \eta =1$, $\gamma = 0.5$, $A=0.01$, $\V=10$, $\mathcal{T}^2=50$ and $\mathrm{Da}=0.01$.}
	\label{vad_fig4}
\end{figure}

\begin{figure}[h!]
	\includegraphics[scale=0.4]{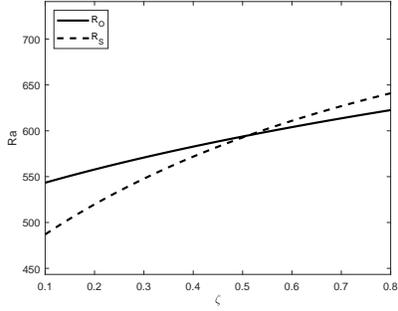}
	\caption{Critical Rayleigh number as function of the solid thermal conductivity parameter $\zeta$ with $\xi= 0.5$, $\eta =1$, $\gamma = 0.5$, $H=100$, $A=0.01$, $\V=10$, $\mathcal{T}^2=50$ and $\mathrm{Da}=0.01$.}
	\label{vad_fig5}
\end{figure}

Figures \ref{vad_fig4a}-\ref{vad_fig4b} show the effect of solid thermal conductivity on the onset of convection. In both Figures, increasing values of $\zeta$ yields growing critical thresholds. This result is physically reasonable since the greater solid thermal conductivity is, the more easily solid matrix absorbs heat from fluid, implying a delay in the onset of convection. Moreover, when $H \rightarrow 0$, the effect of solid thermal conductivity is less remarkable. Assuming that $H$ goes to zero means that heat exchange between fluid and solid is forbidden, that is why changing the solid thermal conductivity parameter $\zeta$ does not affect the critical threshold for the onset of convection. Analogous results were obtained in \cite{Gianfrani2021b}.

\begin{figure}[h!]
	\subfloat[]{
		\includegraphics[scale=0.4]{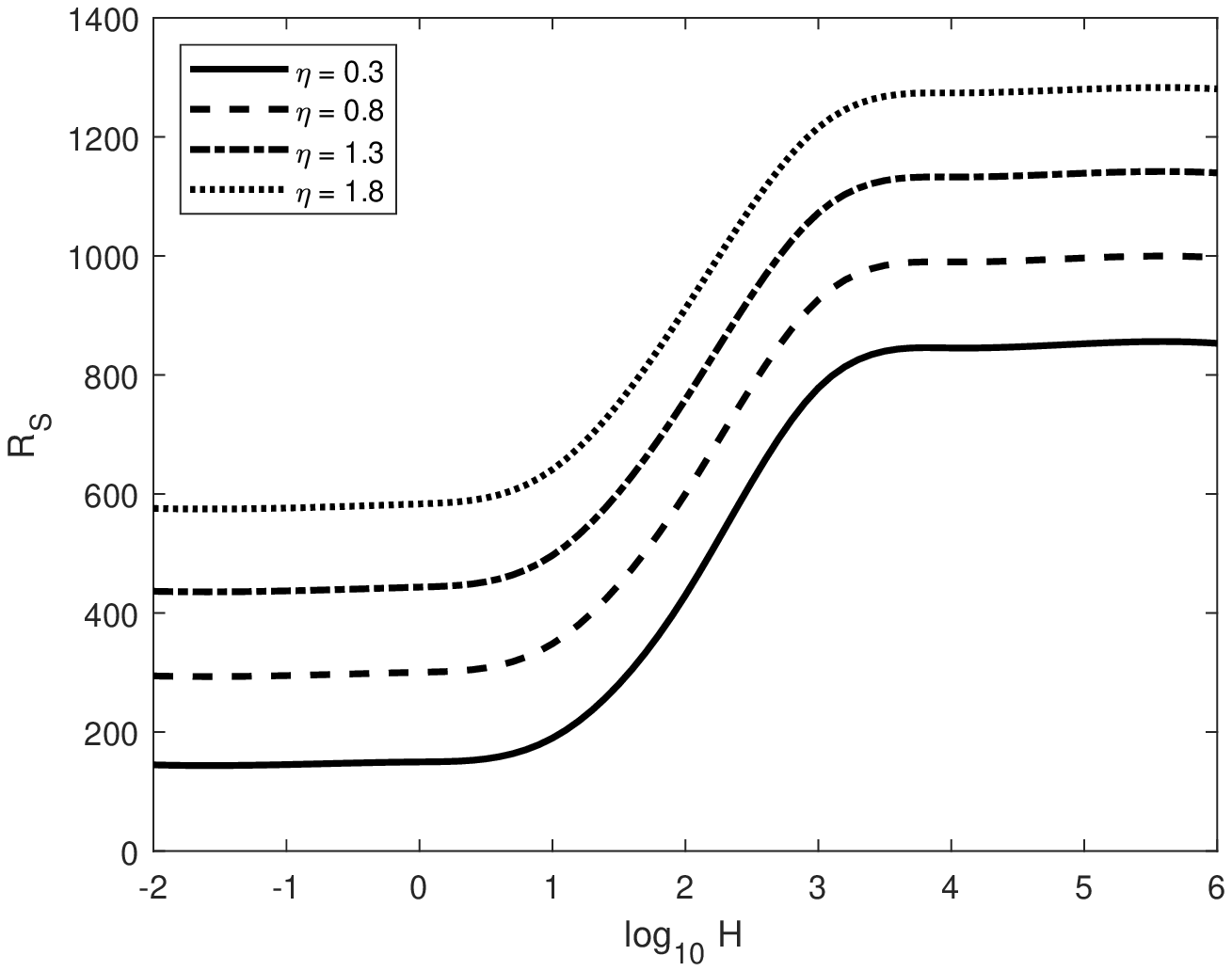}
		\label{vad_fig6a}}
	\subfloat[]{
		\includegraphics[scale=0.4]{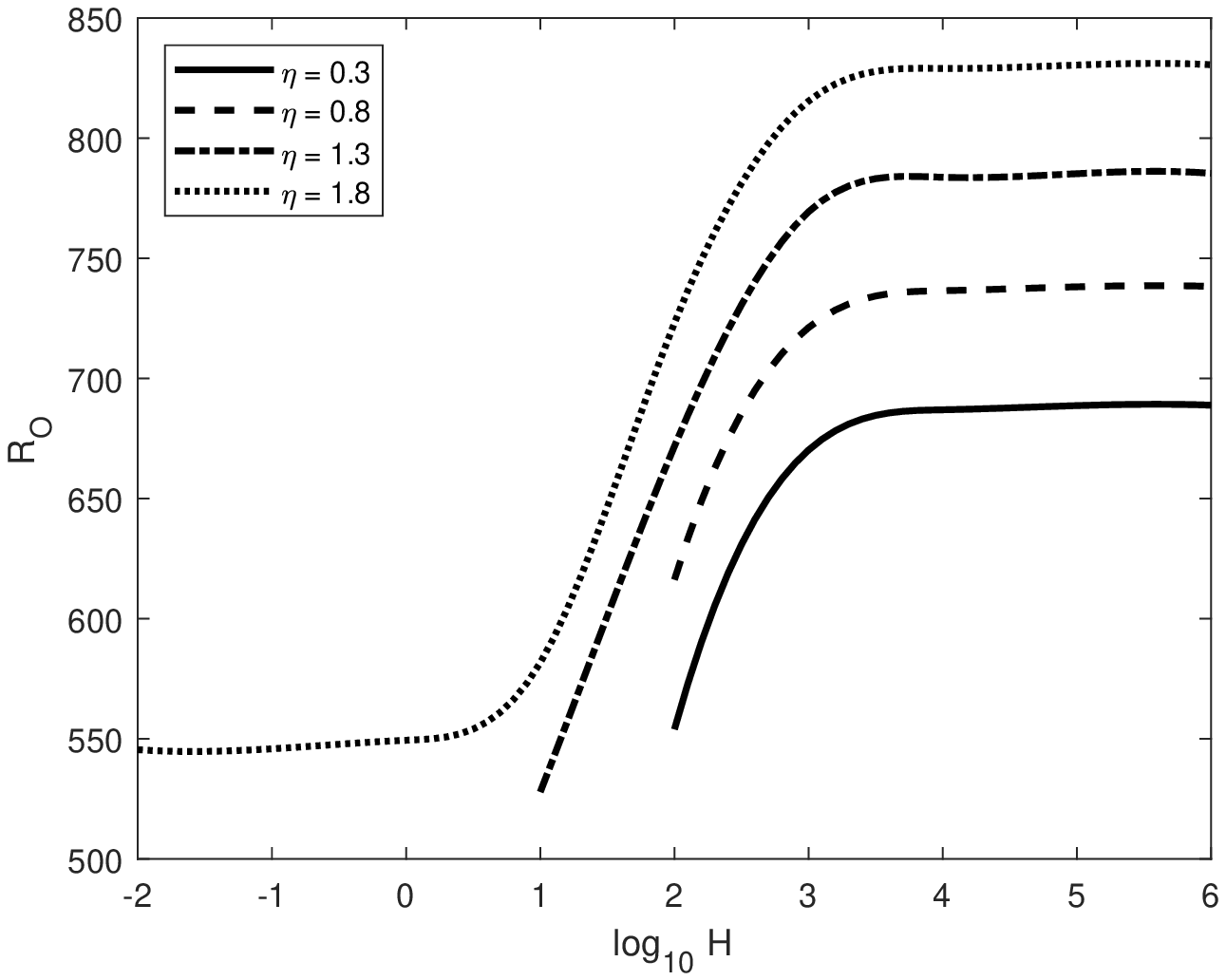}
		\label{vad_fig6b}}
	\caption{Critical Rayleigh number as function of the inter-phase heat transfer coefficient $H$ for different values of the fluid thermal conductivity parameter $\eta$ with $\xi=0.5$, $\zeta =1$, $\gamma = 0.5$, $A=0.01$, $\V=10$, $\mathcal{T}^2=50$ and $\mathrm{Da}=0.01$.}
	\label{vad_fig6}
\end{figure}

\begin{figure}[h!]
	\includegraphics[scale=0.4]{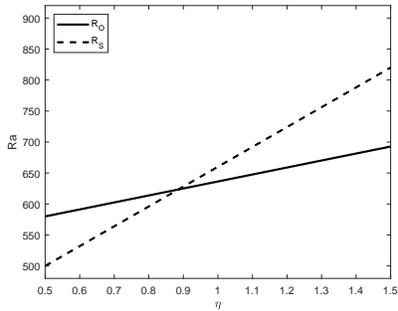}
	\caption{Critical Rayleigh number as function of the solid thermal conductivity parameter $\eta$ with $\xi= 0.5$, $\zeta =1$, $\gamma = 0.5$, $H=100$, $A=0.01$, $\V=10$, $\mathcal{T}^2=50$ and $\mathrm{Da}=0.01$.}
	\label{vad_fig7}
\end{figure}

In Figure \ref{vad_fig5}, the stabilizing effect of $\zeta$ on the onset of instability is evident, as well as the existence of a transition point $\zeta_T$ before which thermal convection occurs through steady motions and beyond which it arises through oscillatory motions.

Figures \ref{vad_fig6a}-\ref{vad_fig6b} show the behaviour of $R_S$ and $R_O$ with respect to variations of the fluid thermal conductivity parameter $\eta$. This parameter has a stabilizing effect on conduction, delaying the onset of convection. Result in Figure \ref{vad_fig6a} is in agreement with the analytical result pointed out in Section \ref{vad_sec2}. From a physical point of view, increasing $\kappa_z^f$, which implies by definition a decreasing $\eta$, allows heat to spread in the vertical direction within the fluid more easily, encouraging the onset of convection. 
We would like to point out that in Figure \ref{vad_fig6b}, conditions (\ref{vad_eq17})-(\ref{vad_eq18}) for the non-existence of oscillatory convection are verified for some values of $H$ and $\eta$, for the set of parameters chosen. As a consequence it is not possible to plot $R_O$ for any $H$ and $\eta$ in Figure \ref{vad_fig6b}.

In Figure \ref{vad_fig7}, it is highlighted  the existence of a threshold value for $\eta$. Once $\eta$ overcomes this critical value, convection occurs through oscillatory motion (i.e. $Ra\equiv R_O$). Before this value, only steady convection can arise ($Ra\equiv R_S$). 

\begin{figure}[h!]
	\subfloat[]{
		\includegraphics[scale=0.4]{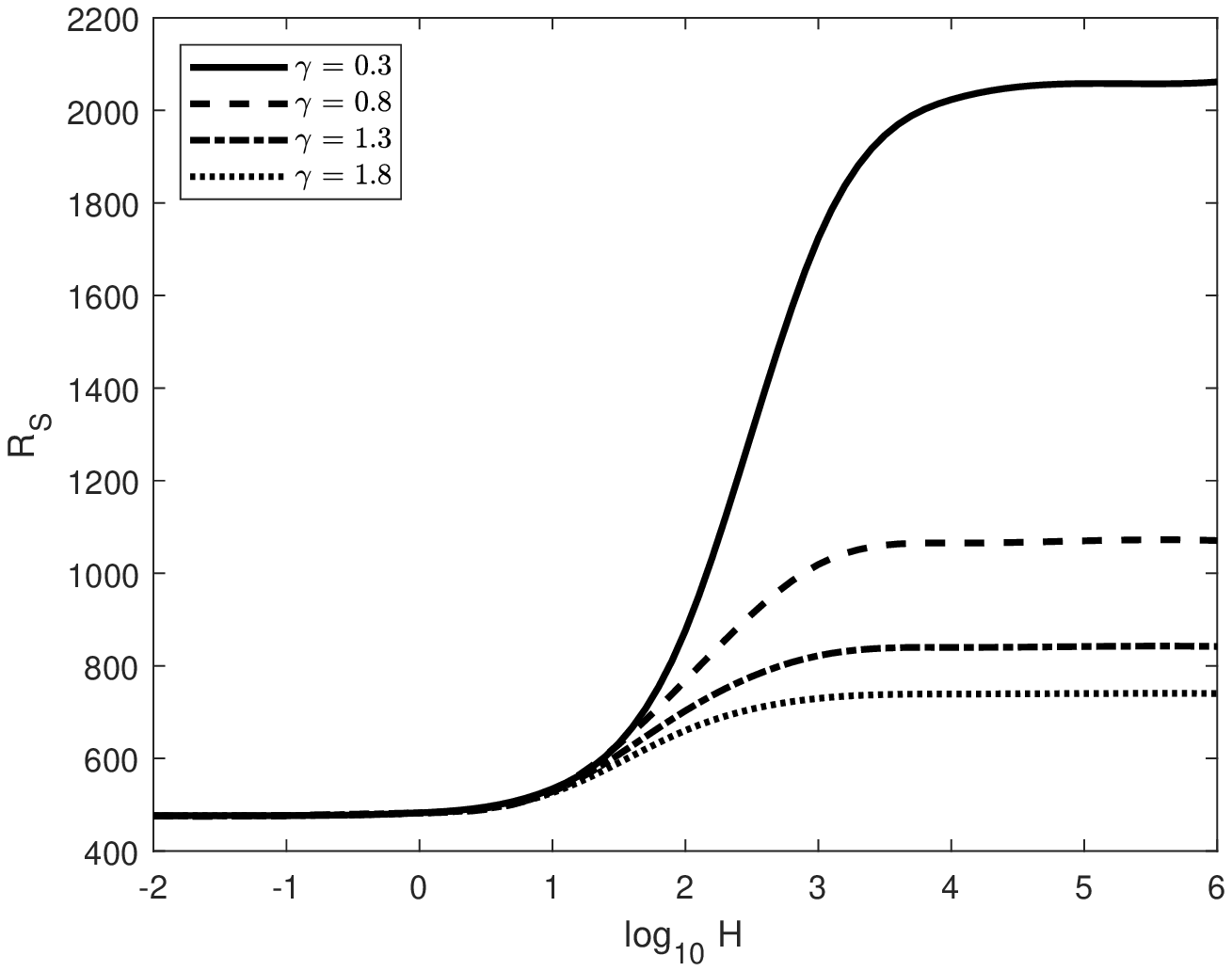}
		\label{vad_fig11a}}
	\subfloat[]{
		\includegraphics[scale=0.4]{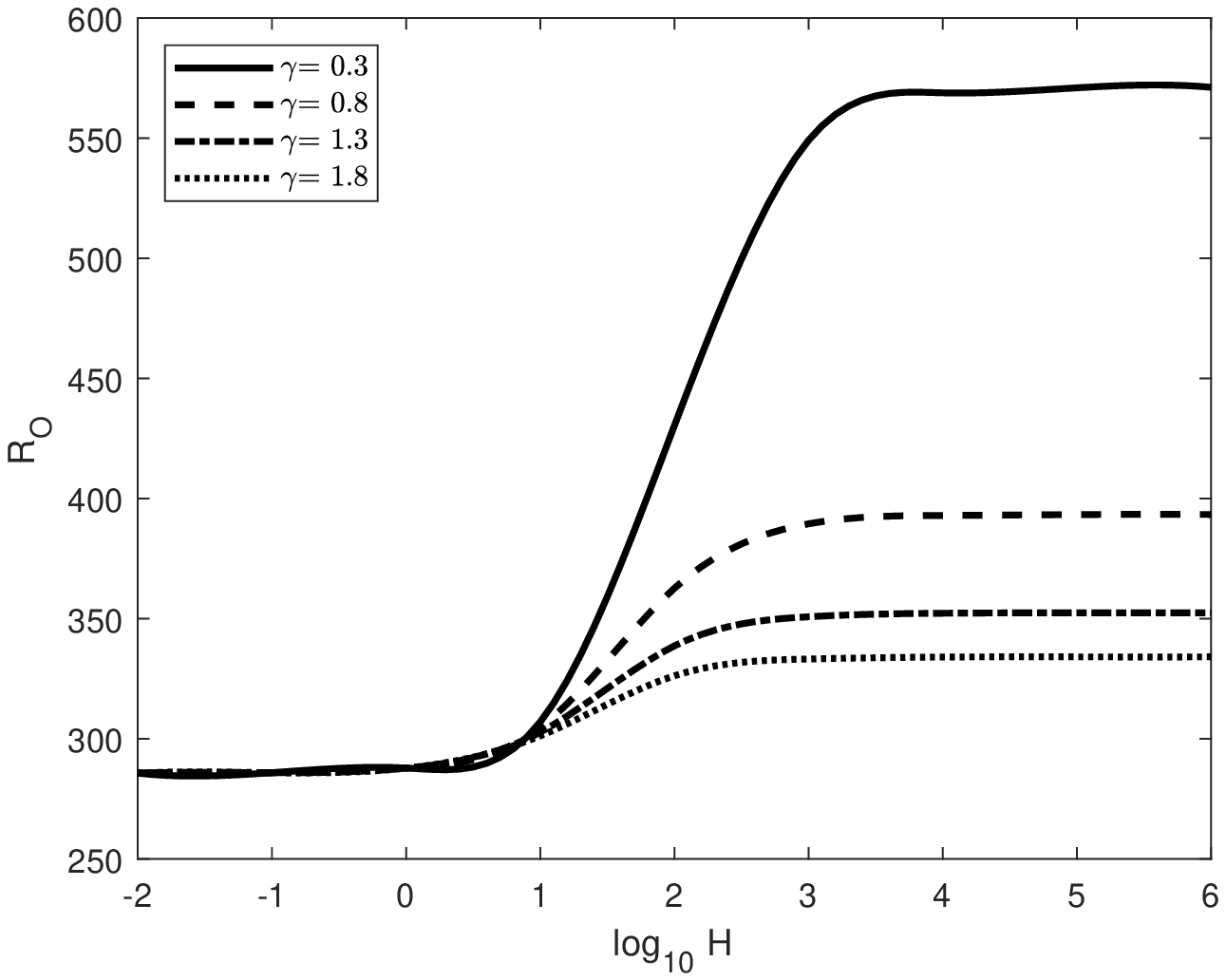}
		\label{vad_fig11b}}
	\caption{Critical Rayleigh number as function of the inter-phase heat transfer coefficient $H$ for different values of the diffusivity ratio $\gamma$ with $\xi=\zeta = \eta =1$, $A=0.01$, $\mathcal{T}^2=50$, $\V=10$ and $\mathrm{Da}=0.01$.}
	\label{vad_fig11}
\end{figure}

The effect of $\kappa_z^f$ on the critical Rayleigh number is highlighted in Figures \ref{vad_fig11a}-\ref{vad_fig11b}, as well. From these Figures, the destabilising effect of $\gamma$ on the critical thresholds comes up and, as consequence, since by definition $\gamma$ is proportional to $\kappa_z^f$, increasing $\kappa_z^f$ encourages the onset of instability.

\begin{table}
	\centering
	\begin{tabular}{lll}
		\hline\noalign{\smallskip}
		$\mathrm{Da}$ & $R_O$ & $R_S$ \\
		\noalign{\smallskip}\hline\noalign{\smallskip}
		0.001	&	\textbf{296.9241}	&	957.2011\\
		0.04	&	\textbf{697.7111}	&	782.9615\\
		0.05	&	796.2326	&	\textbf{783.4380}\\
		0.06	&	895.7849	&	\textbf{786.0607}\\
		0.07	&	996.6127	&	\textbf{790.1571}\\
		0.08	&	$\nexists$	&	\textbf{795.3360}\\
		0.10	&	$\nexists$	&	\textbf{808.0355}\\
		0.20	&	$\nexists$	&	\textbf{896.0463}\\
		0.30	&	$\nexists$	&	\textbf{1004.8}\\
		\noalign{\smallskip}\hline
	\end{tabular}
	\caption{Critical Rayleigh number $Ra$ for different values of $\mathrm{Da}$ with $\xi=\zeta=\eta=1$, $A=0.01$, $\gamma=0.5$, $H=100$, $\V=10$, $\mathcal{T}^2=50$.}
	\label{vad_tab2}
\end{table}

\begin{figure}[h!]
	\includegraphics[scale=0.4]{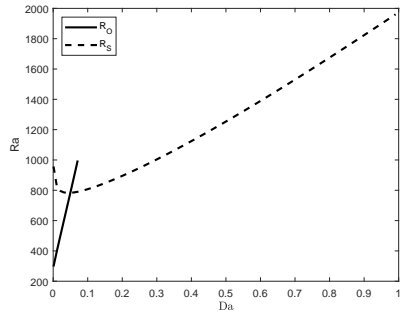}
	\caption{Critical Rayleigh number as function of the Darcy number $\mathrm{Da}$ with $\xi= \eta=\zeta =1$, $\gamma = 0.5$, $H=100$, $A=0.01$, $\V=10$ and $\mathcal{T}^2=50$.}
	\label{vad_fig8}
\end{figure}

In order to capture the influence of $\mathrm{Da}$ on $Ra$ as best as we can, we decided not to plot $R_S$ and $R_O$ as functions of $H$. Instead, we report the behaviour of the critical threshold for a fixed $H$. In Figure \ref{vad_fig8}, a parabolic behaviour of $R_S$ and the existence of a value of $\mathrm{Da}$ beyond which $R_O$ does not exist are evident. Nevertheless, the critical threshold $Ra$ for the onset of instability exhibits an increasing trend with respect to $\mathrm{Da}$, as shown in Table \ref{vad_tab2} where $Ra$ is bold.
When considering the Darcy-Brinkman model, which is closer to a model for clear fluids (in absence of porous medium), we would expect the critical Rayleigh number to get closer to the critical value for clear fluids. It is well known that the critical value for clear fluids is greater than the one for fluids in presence of porous medium (i.e. the presence of a porous medium eases the onset of instability). Hence, result in Figure \ref{vad_fig8} is reasonable.

\begin{figure}[h!]
	\includegraphics[scale=0.4]{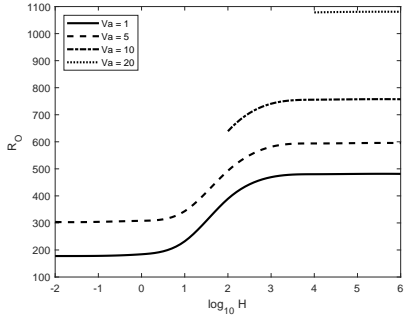}
	\caption{Critical Rayleigh number as function of the inter-phase heat transfer coefficient $H$ for different values of the Vadasz number $\V$ with $\xi=0.5$, $\eta=\zeta=1$, $\gamma = 0.5$, $A=0.01$, $\mathcal{T}^2=50$ and $\mathrm{Da}=0.01$.}
	\label{vad_fig9}
\end{figure}

The influence of the Vadasz number $\V$ on the onset of convection is highlighted in Figure \ref{vad_fig9}. In this case, we report only the behaviour of $R_O$ since, as shown in (\ref{vad_eq13}), $R_S$ does not depend on $\V$. In Figure \ref{vad_fig9}, the stabilising effect of $\V$ is clear. For some values of $\V$ and $H$ it is not possible to plot $R_O$ since conditions for the existence of oscillatory convection are not satisfied.

\begin{figure}[h!]
	\includegraphics[scale=0.4]{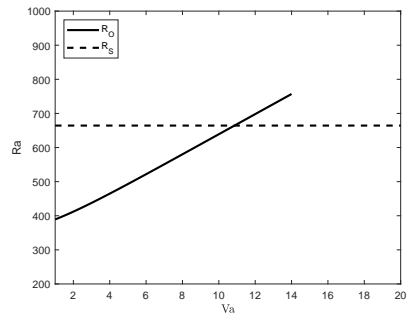}
	\caption{Critical Rayleigh number as function of the Vadasz number $\V$ with $\xi=0.5$, $\eta=\zeta=1$, $\gamma = 0.5$, $H=100$, $A=0.01$, $\V=10$, $\mathcal{T}^2=50$ and $\mathrm{Da}=0.01$.}
	\label{vad_fig10}
\end{figure}

A focus on the behaviour of $Ra$ with respect to $\V$, for a fixed $H$, is given in Figure \ref{vad_fig10}, where it is shown that increasing Vadasz number makes the critical threshold increase at least up to a certain value, beyond which the critical Rayleigh number is constant and convection arises through steady motion. If $\V \rightarrow \infty$, as we have pointed out in Remark \ref{vad_rem2}, oscillatory convection cannot occur. In fact, looking at model (\ref{vad_modad}), when $\V \rightarrow \infty$, the inertia term disappears and model (\ref{vad_modad}) reduces to that one studied in \cite{Gianfrani2021b} for which the principle of exchange of stabilities holds and convection occurs only through steady motion. On the other hand, if $\V \rightarrow 0$, the inertia term strongly affects the model and instability occurs earlier, through oscillatory motion. As a result, the inertia term encourages the onset of convection and such a result is not surprising. Indeed, we can write by definition $\V=\dfrac{\mathrm{Pr}}{\mathrm{Da} \ c_a}$, where $\mathrm{Pr}=\dfrac{\tilde\mu c_f}{\kappa_z^f}$ is the Prandtl number and $c_a$ is inversely proportional to $\varepsilon$ \cite{Nield-Bejan}. Then, it immediately follows that if porosity $\varepsilon \rightarrow 0$, i.e. the medium becomes less porous, then $\V \rightarrow 0$ and the critical Rayleigh number decreases, which is expected as the presence of a porous medium has a destabilising effect on conduction, as already pointed out.

\section{Conclusions}
The study undertaken in this paper was devoted to investigate the effect of the inertia term in the momentum equation of a Darcy-Brinkman model on the onset of convective motions. In this paper, we performed a linear instability analysis to show that the presence of the Vadasz term leads to different physical phenomena. Specifically, the presence of inertia term makes the onset of convection possible via either oscillatory or steady motions. Moreover, we proved analytically that the Vadasz term does not affect steady convection, namely we recovered same results as in \cite{Gianfrani2021b}. Conditions for the non-existence of oscillatory convective motions were determined numerically because of the high complexity of the critical Rayleigh number expression.

In addition, we studied the effect of parameters on the onset of both steady and oscillatory convective motions. We proved analytically that the influence of anisotropic permeability $\xi$ on steady convection depends on the Taylor number $\mathcal{T}$, while numerically we showed the stabilising effect of rotation, thermal conductivities (both fluid and solid one) and of the Darcy number $\mathrm{Da}$ on the onset of instability.

\bibliographystyle{ieeetr}
\bibliography{mybib}
\end{document}